\documentstyle[12pt]{article}

\begin{document}
\pagestyle{plain}

\newcommand{\be}{\begin{equation}}
\newcommand{\ee}{\end{equation}}
\newcommand{\bea}{\begin{eqnarray}}
\newcommand{\eea}{\end{eqnarray}}
\newcommand{\vp}{\varphi}
\newcommand{\pr}{\prime}

\title{Solitary Waves and Compactons \\ in a class of Generalized
Korteweg-DeVries Equations}
\author{
Fred Cooper \\
{\small \sl Theoretical Division, Los Alamos National Laboratory,}\\
{\small \sl Los Alamos, NM 87545}\\
{\small \sl and   }\\
{\small \sl Physics Department, University of New Hampshire,} \\
{\small \sl Durham, NH 03824}\\
\and Harvey Shepard \\
{\small \sl Physics Department, University of New Hampshire,} \\
{\small \sl Durham, NH 03824}\\
\and Pasquale Sodano \\
{\small \sl Dipartimento di Fisica and Sezione, INFN Universita di Perugia,} \\
{\small \sl 06100 Perugia, Italy}}
\maketitle

\begin{abstract}
We study the class of generalized Korteweg-DeVries equations derivable
from the Lagrangian:
$ L(l,p) = \int \left( \frac{1}{2} \vp_{x} \vp_{t} - { {(\vp_{x})^{l}} \over
{l(l-1)}} + \alpha(\vp_{x})^{p} (\vp_{xx})^{2}  \right) dx, $
where the usual fields $u(x,t)$ of the generalized KdV equation
are defined by $u(x,t) = \vp_{x}(x,t)$.
This class contains compactons,  which
are solitary waves with compact support, and when $l=p+2$, these
solutions have the feature that their width  is independent of the
amplitude. We consider the Hamiltonian structure and integrability properties
of this class of KdV equations. We show that many of the
properties of the solitary waves and compactons are easily obtained using
a variational method based on the principle of least action. Using a class
of  trial variational functions of the form $u(x,t) =  A(t) \exp
\left[-\beta (t) \left|x-q(t)  \right|^{2n} \right]$ we find
soliton-like  solutions for all $n$,
moving with fixed shape and constant velocity, $c$. We show that the velocity,
mass, and energy of the variational travelling wave solutions are related by  $
c = 2 r E M^{-1}$, where $ r = (p+l+2)/(p+6-l)$, independent of $n$.\newline

\end{abstract}

\section{Introduction}

Recently, Rosenau and Hyman \cite{RH} have shown that in a particular
generalization of the KdV equation, defined by parameters (m,n), namely

\be
 K(n,m) :   u_t + (u^m)_x + (u^n)_{xxx} = 0,
\ee
that a new form of solitary wave with compact support  and width
independent  of amplitude exists.  For their choice of generalized KdV
equations the compactons with $m=n \le{3}$ had the form $[cos(a
\xi)]^{2/(m-1)}$, where $\xi= x-ct$
and for m=2,3 they obtained:

\bea
    K(2,2) &  &  u_c = { {4c} \over {3}} \cos^2(\xi/4) \nonumber\\
    K(3,3) &  &  u_c = \left({3c} \over {2} \right) ^{1/2} \cos(\xi/3) .
\eea

Unlike the ordinary KdV equation, the generalized KdV equation
considered by Rosenau and Hyman  was not derivable from a first
order Lagrangian except for $n=1$, and  did not possess
the usual conservation laws of energy
 and mass that the KdV equation possessed.  It is presumed that the
generalized  KdV equations found by the above authors are not
completely integrable, but instead possess only a finite number of
conservation laws.
   Because of this, we were led to  consider a
different  generalization of the KdV equation based on a first
order Lagrangian formulation.  That is, we
consider

\be
L(l,p) = \int \left( \frac{1}{2} \vp_{x} \vp_{t} - { {(\vp_{x})^{l}} \over
{l(l-1)}} + \alpha
(\vp_{x})^{p} (\vp_{xx})^{2}  \right) dx. \label{L}
\ee

This Lagrangian leads to a generalized  sequence of KdV equations of the
form:
\be
K^*(l,p):  u_t= u_x u^{l-2} + \alpha\left(2 u_{xxx}u^p + 4p u^{p-1} u_x
u_{xx} +p(p-1) u^{p-2} (u_x)^3 \right)\label{kstar}
\ee
where
\be
u(x) = \vp_x(x) .\nonumber
\ee
These equations have the same terms as the equations considered by
Rosenau and Hyman, but the relative weights of the terms are
quite different leading to the possibility that the integrability
properties might be different. [For the purposes of comparison it may be
helpful to note that their set $(m,n)$ corresponds to our $(l-1, p+1)$.]The
rest of the paper is organized
follows:  In section 2 we discuss some exact travelling wave solutions to
(\ref{kstar}).
In section 3 we derive the conservation laws and discuss the
Hamiltonian structure of these equations. In section 4 we apply
the time dependent variational approach to obtaining approximate
solitary wave solutions, and in section 5 we compare the variational
solutions to the exact ones.
\section{Exact solitary wave and compacton solutions}
If we assume a solution to (\ref{kstar}) in the form of a travelling
wave:
 \be
u(x,t) =f(\xi) =f(x+ct) ,
\ee
one obtains for f:
\be
 c f^{\prime} = f^{\prime} f^{l-2} + \alpha\left(2 f^{\prime \pr \pr}
f^p + 4p f^{p-1} f^{\prime} f^{\prime \pr} +p(p-1) f^{p-2} f^{\prime
3}\right). \ee
Integrating twice we obtain:
\be
   {{c} \over {2}} f^2 -{{f^l} \over {l(l-1)}} - \alpha f^{\prime 2}f^p=
C_1 f + C_2.
\ee
We seek solutions where the integration constants, $C_1$ and $C_2$ are zero.
This puts lower bounds on $l$ and $p$: $l > 1$ and $ f^{\prime \pr }
f^p \rightarrow 0,  f^{\prime 2}f^{p-1} \rightarrow 0$ at edges where $f
\rightarrow 0$.
Then we obtain
\be
\alpha f^{\pr 2} =  {{c} \over {2}} f^{2-p} -{{f^{l-p}} \over {l(l-1)}}.
\ee
For finite $f^{\prime}$ at the edges, we must have $p\leq 2, l\geq p$.

Let us now look at some special examples. (Note that we have chosen signs so
that all travelling waves have $u>0$ and move to the left.)
The usual KdV equation has $\alpha=1/2$,
$l=3$, $p=0$.
For that case one has the well known soliton:
\be
u= (3 c ) \rm{sech}^2 \left[ \sqrt{3 c/2} \xi \right].
\ee
We define the ``mass'' M via
\be
M= \int_{-\infty} ^{\infty}  dx  [u(x,t)]^2 .\nonumber
\ee
For this solution we find that we can express $M$ and $E$ in terms of
$c$  as follows: $ M=  24 c^{3/2}  $,  $ E =  {{36} \over {5}} c^{5/2}$
so that \be
c = {{10}\over {3}}  E M^{-1} =(M/24)^{2/3}.
\ee

The case $l=p+2$ is the case relevant for compactons whose
width is independent of  the velocity c.
For $p=1$, $\alpha=1/2$  one obtains the compacton solution:
\be
u_1 = 3c \cos^2(\xi/\sqrt{12})  \label{c1},
\ee
where $  |\xi | \le \sqrt{3} \pi$.
One finds:
$ M=  {{27} \over {4}}\pi \sqrt{3} c^{2}  $,  $ E = {{27} \over {8}}
\sqrt{3} \pi c^3  $
 so that
\be
c =   2  EM^{-1} = \left( {{4M} \over {27\pi\sqrt{3}}}\right)^{1/2}.
\ee
There is another compacton solution with
 $p=2$,  $\alpha=3$. \be
u_2 = \sqrt{6 c} \cos(\xi/6)
\ee
with $  |\xi | \le 3 \pi$.
For this compacton,  one finds
$ M=  18 \pi c  $,  $ E = {{9 \pi c^2}
\over {2}} $
 so that
\be
c =   4 E M^{-1}= {{M} \over {18 \pi}}.
\ee

For the values, $l=3$, $p=2$  there is  a compacton whose  width
depends on the velocity . Choosing $\alpha = 1/4$ we find:
\be
u = 3c  -(\xi^2)/6  \label{parab}
\ee
on the interval
\be
|\xi| \leq 3 \sqrt{2c};
\ee
elsewhere it is zero.
For this compacton one finds:
$M=  {{144} \over {5}} \sqrt{2}c^{5/2}$,  $E= {{72}\over{7}} \sqrt{2}
c^{7/2} $ so
\be
c  =    { 14 \over 5}  E M^{-1} = \left({{5M} \over {144
\sqrt{2}}}\right)^{2/5}.
 \ee

Thus, apart from constants we find the same functional form
for the compactons for our generalized KdV equations as those found
by   Rosenau and Hyman in their different generalization of the KdV
equation.

\section{Conservation laws and canonical structure}

Equation (\ref{kstar}) can be written in canonical form displaying the
same Poisson bracket structure as found for the KdV equation:

\be
u_t = \partial_x  {{\delta H } \over {\delta u}}
= \{u,H\}
\ee
where H is the Hamiltonian obtained  from the Lagrangian (\ref{L}),

\bea
H &=& \int \left[ (\pi \dot{\vp}) - L \right] dx \nonumber \\
 &=& \int \left[ { {(\vp_{x})^l} \over {l(l-1)}} -\alpha
(\vp_x)^p (\vp_{xx})^{2} \right] dx, \nonumber \\
&=& \int \left[ { {u^l} \over {l(l-1)}} -\alpha
u^p (u_{x})^{2} \right] dx.\\
  \label{Hamiltonian}
\eea

By the usual arguments \cite{Das} this is consistent with a Poisson
bracket structure
\be
\{u(x),u(y)\} = \partial_x \delta (x-y) .\label{poiss1}
\ee
Let us now show that we have a system of equations which have exactly
the  same first three conservation laws as the ordinary KdV equation,
namely the area, mass and energy. This is unlike the
equations studied by Rosenau and Hyman that did not conserve the
mass and energy, but instead had different conserved quantities.

We have
\be
u_t = \partial_x  {{\delta H } \over {\delta u}}
\ee
so that the ``area'' under $u(x,t)$ is conserved:
\be
    l\int u(x,t) dx \equiv H_0 \label{h0}
\ee

Multiplying by $u(x,t)$ we find:

\be
\partial_t({{u^2} \over {2}} ) = \partial_x \left[ {{u^l} \over {l}
}+\alpha \{(p-1) u^p u_x^2 + 2 u^{p+1} u_{xx} \} \right]
\ee

which leads to the conservation of ``mass''
\be
(1/2) \int u^2(x,t) dx = (1/2) M \equiv   H_1  \label{h1}
\ee
For the KdV equation $H_1$ was a second Hamiltonian
under a second Poisson bracket structure.
{}From Lagrange's equations we immediately get a third
conservation law, the energy:
\be
H =  \int \left[ { {u^l} \over {l(l-1)}} -\alpha
u^p (u_{x})^{2} \right] dx \equiv H_2.
\ee
The energy provided the first Poisson bracket structure:
Considering the mass as a second Hamiltonian, the
KdV equation has a second Poisson bracket structure using
$H_1$.
Assuming
\be
u_t = \{u,H_1 \} = \int dy \{u(x),u(y)\}_1 {{\delta H_1 } \over {\delta u(y)}}
,
\ee
one finds for the KdV equation  that
\be
\{u(x), u(y)\}_1= \left( D^3 +{{1}\over{3}} (Du+uD) \right) \delta(x-y)
\ee
where $ D= \partial_x$.
With this assumed Poisson bracket structure one again recovers the KdV
equation. This Poisson bracket structure is identical to the Virasoro
algebra with a specific central charge.  This fact enables one to
show that there is an infinite number of conservation laws in the KdV
equation, and it is an exactly integrable system \cite{Das}.

For the generalized KdV equations
we find that we can   write
\be
u_t =\left( \alpha ( D^2 u^p D + D u^p D^2) +
{{1}\over {l}}  (Du^{l-2} +u^{l-2} D) \right) u
\ee
so that there is a chance for a second Hamiltonian if the
Jacobi identity is satisfied.  One can postulate that the
second Poisson bracket structure is given by
\be
\{u(x), u(y)\}_1 = \left( \alpha ( D^2 u^p D + D u^p D^2) +
{{1}\over {l}}  (Du^{l-2} +u^{l-2} D) \right) \delta (x-y)
\label{poiss2} .
\ee

So we need to show for what $l,p$ this bracket structure
obeys the Jacobi Identity,where   the bracket is defined by :
\be
\{F[u],G[u] \} = \int_{-\infty} ^{\infty} dx dy{\delta F \over \delta
u(x)} \{u(x),u(y)\}_1 {\delta G \over \delta u(y)}.
\ee
One can show immediately that the Hamiltonians $H_1$ and $H_2$ commute
using either Poisson Bracket structure (\ref{poiss1}) or (\ref{poiss2}).

We have that
\be
\{H_2[u],H_1[u] \} = \int_{-\infty} ^{\infty} dx dy{\delta H_2 \over
\delta u(x)} \{u(x),u(y)\}_1 {\delta H_1 \over \delta u(y)}
\label{comm}. \ee

For the usual bracket structure (\ref{poiss1}) we can rewrite
(\ref{comm}) as
\bea
\{H_2[u],H_1[u] \}  &=&\int_{-\infty} ^{\infty} dx u_t(x)
\left( {\delta H_1 \over
\delta u(x)}\right) \nonumber \\
&=& {1 \over 2} \partial_t \int_{-\infty} ^{\infty} dx u^2(x,t) =0.
\eea

For the second bracket structure (\ref{poiss2}) we have
instead:
\bea
\{H_2[u],H_1[u] \}_1  &=&\int_{-\infty} ^{\infty} dx u_t(x)
\left( {\delta H_2 \over
\delta u(x)}\right) \nonumber \\
&=&\int_{-\infty} ^{\infty} dx  {\delta H_2 \over
\delta u(x)} \partial_x \left( {\delta H_2 \over
\delta u(x)}\right)
\nonumber \\
&=& {1 \over 2} \int_{-\infty} ^{\infty} dx \partial_x \left( {\delta H_2 \over
\delta u(x)}\right)^2 =0.
\eea

Encouraged by this result we have attempted to repeat the induction
proof of the existence of an infinite number of
conservation laws, assuming as in the KdV equation that one has the
conservation laws obey the recursion relations:

\be
\left( \alpha ( D^2 u^p D + D u^p D^2) +
{{1}\over {l}}  (Du^{l-2} +u^{l-2} D) \right) {{\delta H_{n-1}} \over
{\delta u(x)}} =
D  {{\delta H_{n}} \over
{\delta u(x)}}.
\ee
Starting with $H_0$ defined
by (\ref{h0}) we get the candidate Hamiltonian:

\be
H_1 =
\int { {u^{l-1}(x,t)} \over {(l-1)}} dx  \label{h1new}
\ee
instead of (\ref{h1}).  If we now ask if this is conserved by
considering the equation
\be
{{dH_1} \over {dt}} = \{H_1,H_2 \} \label{cons}
\ee
using the first Poisson bracket structure, we find that the right
hand side of (\ref{cons}) is not a total divergence
unless $l=3$.  For $l=3$ one has :
 \be
\left( \alpha ( D^2 u^p D + D u^p D^2) +
{{1}\over {l}}  (Du^{l-2} +u^{l-2} D) \right) {{\delta H_{1}} \over
{\delta u(x)}} =
D {{\delta H_{2}} \over
{\delta u(x)}}
\ee

  However if we iterate one more time (with $l=3$)
we obtain:
\be
\left( \alpha ( D^2 u^p D + D u^p D^2) +
{{1}\over {3}}  (Du +u D) \right) {{\delta H_{2}} \over
{\delta u(x)}} =
D  F_3 (x)
\ee
and we find by explicit construction that $F_3 (x)$ is not the
variational derivative of a local Hamiltonian unless $p=0$. Thus
this bi-Hamiltonian method of finding an infinite number of conservation laws
only
works for  the original KdV equation.
 We  surmise that (\ref{poiss2}) is
not a
 valid bracket structure and that

\be
\{H[u],\{F[u],G[u]\}\} + \{G[u],\{H[u],F[u]\}\}+\{F[u],\{G[u],H[u]\}\}=0
\ee
is {\it not} satisfied for the postulated second bracket. Thus we have not
succeeded in showing that these new equations are exactly integrable, and we
are in the same situation, in spite of having a first order
Lagrangian, as for the generalized KdV equations of Rosenau and Hyman
\cite{RH}.

We have not as yet performed numerical simulations of the scattering of our new
compacton solutions. For Rosenau and Hyman such numerical experiments produced
behavior very similar to but not exactly the same as that observed in
completely integrable systems, namely, stability and preservation of shape.
They find that elastic collisions are accompanied by the production of low
amplitude compacton-anticompacton pairs \cite{RH}.

\section {Variational approach}

Our time-dependent variational approach for studying solitary waves
is related to Dirac's variational approach to the Schrodinger
equation \cite{Dirac}, \cite{Jackiw-Kerman}.    In our previous work
\cite{CSLS1} \cite{CSLS2}, we introduced a post-Gaussian variational
approximation, a continuous family of trial variational functions more general
than  Gaussians, which can still be treated analytically.
Assuming a variational ansatz of the  form $u(x,t) =  A(t) \exp
\left[-\beta (t) \left| x-q(t) \right|^{2n}  \right]$,we will extremize
the effective action for the trial wave functional and determine the
classical dynamics for the variational parameters.  We will find that
for all (l,p) the dynamics of the variational parameters lead to
solitary waves moving with constant velocity and constant amplitude.  For the
special case of $l=p+2$ we find immediately that the width of the
soliton is independent of the amplitude and velocity. Correct functional
relations between energy, mass, amplitude and velocity are obtained very
quickly from the variational method, although one does not find that
the  $l=p+2$ variational solitons have compact support. We will find that most
of the properties  of the single ``soliton'' solutions to these equations
can be  obtained by using this very simple trial wave function ansatz
and extremizing the action.

The starting point for the variational calculation is the action

\be
\Gamma = \int L dt, \label{action}
\ee
where $L$ is given by (\ref{L}).

Just as we did in our study of the KdV equation
we choose a trial wave function of the form:

\be
u_{v}(x,t) = A(t) \exp \left[-\beta(t) |x-q(t)|^{2n} \right],
\label{uv} \ee
where $n$ is an arbitrary continuous, real parameter.

The variational parameters have a simple interpretation in terms of expectation
values with respect to the ``probability''
\be
P(x,t) = \frac{\left[u_{v}(x,t)\right]^{2}}{M(t)},
\ee
where the mass $M$ is defined as above

\be
M(t)\equiv \int \left[u_{v}(x,t)\right]^{2} dx \label{M}.
\ee
(Here we allow $M$ to be a function of $t$, even though $M$ is
conserved)

Since $\langle x-q(t) \rangle = 0$, $q(t) = \langle x \rangle$. From
(\ref{M}) and (\ref{uv}) we have

\be
A(t) =  \frac{M^{1/2} (2\beta)^{1/4n}}{\left[2\Gamma\left(\frac{1}{2n}
+  1\right)\right]^{1/2}}.
\ee
The inverse width $\beta$ is related to

\be
G_{2n} \equiv \langle |x-q(t)|^{2n} \rangle = \frac{1}{4n\beta}.
\ee

Following our approach in \cite{CSLS2}, we find that the action for the
trial wave function (\ref{uv})  is given by:

\bea
\Gamma(q,\beta,M,n) & = & \int \left( -\frac{1}{2} M \dot{q} - C_{1}(n)
\beta^{(l-2)/4n}
M^{l/2} + C_{2}(n) M^{1+p/2}  \beta^{(p+4)/4n} \right) dt \nonumber \\
& \equiv & \int L_{1} (q,\dot{q},M,\beta) dt,
\eea
where

\bea
C_{1}(n) & = & {{1} \over {l(l-1)}}
\left(\frac{2^l}{l^2}\right)^{1/4n} \left[ 2\Gamma\left(\frac{1}{2n} +
1\right) \right]^{(2-l)/2} \nonumber \\ \\
C_{2}(n) & = &  4\alpha  n (2)^{(p+2)/4n}(2+p)^{{{1} \over{2n}}
-2} \frac{\Gamma\left(2 -\frac{1}{2n} \right)}
{[2\Gamma\left(\frac{1}{2n} + 1\right)]^{1+p/2}}. \nonumber \eea
 We eliminate the variable of constraint $\beta$ (using
$\delta\Gamma/ \delta\beta = 0$) and  find

\be
\beta  =\left[d(n)\right]^{4n} M^{2n(p+2-l)/(l-p-6)} \label{beta},
\ee
where
\be
d(n) = \left[{{(p+4) C_{2}(n)}\over
{(l-2)C_{1}(n)}}\right]^{1/(l-p-6)}.\label{d}
\ee

{}From (\ref{beta})  we see that when
\be
l=p+2,
\ee
 the width of the soliton $\beta$  does not depend on M and thus is
independent of the amplitude or velocity.  This special case is
precisely the case when the exact  solution is a compacton.

We now eliminate $\beta$ in favor of $M$, and symmetrizing the
Lagrangian (\ref{L})   we obtain \cite{Jackiw-Faddeev}:

\be
L_{2} = \frac{1}{4} \left(q\dot{M} - \dot{q}M \right) - H(M),
\label{L_2}
\ee
where

\be
H(M) = \left( C_{1} d^{(l-2)} - C_{2} d^{(p+4)} \right) M^{r},
\ee
where
$ r = (p+l+2)/(p+6-l)$.
 Extremizing the action yields:

\bea
\dot{M}  =  0 & \Longrightarrow& M  =  {\rm const.} \nonumber \\
& \Longrightarrow& \beta = {\rm const.}
\eea
and

\be
\dot{q} = -2r \left( C_{1} d^{(l-2)} - C_{2} d^{(p+4)} \right) M^{r-1},
\ee
as well as a conserved energy

\be
E = \left( C_{1} d^{(l-2)} - C_{2} d^{(p+4)} \right) M^{r}  \label{E}
\ee
Thus the velocity of the solitary wave is constant and can be related to
the conserved energy via

\be
\dot{q} = -c = -2 r E M^{-1}. \label{q-dot}
\ee

This is precisely the form we obtained for the exact solutions.

We have not yet extremized the action with respect to the variational
parameter $n$  which is equivalent to extremizing the energy with
respect to $n$. We perform this extremization graphically
for each value of $l, p $.  The explicit
form of the trial wave function is:
 \bea
u_{v}(x,t)& =& d[n,p,l]  M^{2/(p+6-l)} 2^{1/4n}  \left[
2\Gamma\left(1+\frac{1}{2n}  \right) \right]^{-1/2}\times \nonumber \\
&&\mbox{} \exp \left [-d^{4n} M^{2n(l-p-2)/(p+6-l)}
|x+ct-x_{0}|^{2n}\right],  \label{uvex}
\eea
where d is given by (\ref{d}).

Now let us see how these trial wave functions compare with the
exact answers for special cases. Since we explicitly know the
M dependence of the answer, we can set M=1 as our normalization for
both the variational and exact solitons.

First let us review the results for the KdV equation:
Here the variational wave function is obtained by first
setting $p=0,\alpha=1/2$ and $l=3$.  One then extremizes the action
in the constant parameter n.  We find numerically that
n=.877, which also extremizes the energy to be
$.035999 M^{5/3}$ and determines the velocity to be $.119995 M^{2/3}$.
In figure 1a and 1b we compare this variational result to the
exact soliton given by:

\be
u= (3 c)  \rm{sech}^2\left[ {\sqrt{3c/2}}(x+ct)\right].
\ee
We see from fig 1b that globally we achieve an accuracy of better than
$1\%$.
For this solution the velocity and energy are:

\be
c =  (M/24)^{2/3}= .120187 M^{2/3}
\ee
\be
E= {3 \over10} M c = .0360562 M^{5/3}.
\ee
Thus we obtain the velocity (and the energy) accurate to $0.2 \% $ from the
variational calculation.

Next let us look at the compacton that is a segment of a parabola
(\ref{parab}).
Here the variational wave function is obtained by
setting $p=2,\alpha=1/4$ and $l=3$.  Mininimizing the action
in the constant parameter n  we find numerically that
n=1.423, which also extremizes the energy to be
.0803831 and determines the velocity to be .225073
In figure 2a and 2b we compare this variational result to the
exact soliton given by:
\be
u = 3c  -{{(x+ct)^2} \over {6}}
\ee
on the interval
\be
|\xi| \leq 3 \sqrt{2c}
\ee
otherwise zero.
We notice that the global accuracy is a few per cent except near
the place where the true compacton goes to zero.
For this compacton one finds:
\be
c  =   \left({{5M} \over {144
\sqrt{2}}}\right)^{2/5} = .227006 M^{2/5}
 \ee
and
\be
E=  5/14 Mc =     .0810735 M^{7/5}
\ee
Thus we find that the velocity (and the energy) are determined
to $0.8\% $ accuracy.

Next let us look at the compacton given by (\ref{c1}).
Here the variational wave function is obtained by
setting $p=1,\alpha=1/2$ and $l=3$.  Extremizing the action
in the constant parameter n  we find numerically that
n=1.154, which also extremizes the energy to be
.054888 and determines the velocity to be .164666
In figure 3a and 3b we compare this variational result to the
exact soliton given by:

\be
u_1 =  3c \cos^2(\xi/\sqrt{12}),
\ee
where $ \xi \le \sqrt{3} \pi$

We notice that the global accuracy is a few per cent except near
the place where the true compacton goes to zero.
For this compacton one finds:

\be
c  = \left( {{4M} \over {27\pi\sqrt{3}}}\right)^{1/2}= .165003 M^{1/2}
\ee
For the exact solitary wave one has the relationship:
\be
E= cM/2 =.0825017 M
\ee

The approximate soliton had instead:
\be
E= cM/(2r) = cM/3 = .054888M
\ee
So for this compacton, the variational energy is wrong by a factor
of $2/3$ although the velocity is correct to $0.2\%$.
This is the only case where the variational method does not give the
exact relationship between energy and velocity. However, we note that with a
change in the sign of the second term in the expression for E, eq. (\ref{E}),
the variational energy becomes $.0823332 M$, which is accurate to $0.2\%$,
leaving the velocity and optimal n unchanged! (Such a sign change would follow
from a factor $(-1)^p$ in $C_2$, which would leave all other results in this
work unchanged.)

Finally let us look at the compacton given by (15).
Here the variational wave function is obtained by
setting $p=2,\alpha=3$ and $l=4$.  Extremizing the action
in the constant parameter n  we find numerically that
n=1.283, which also extremizes the energy to be
.00436284 and determines the velocity to be .017451
In figure 4a and 4b we compare this variational reult to the
exact soliton given by:

\be
u_2 = \sqrt{6 c} \cos(\xi/6)
.\ee
where $ \xi \le 3 \pi$

We notice that the global accuracy is a few per cent except near
the place where the true compacton goes to zero.
For the exact compacton one finds:

\be
c  = {M \over {18 \pi}}= .0176839 M
\ee
The variational estimates for c and E are accurate to $ 1.3\% $. For the exact
and variational soliton one has the same relationship:
\be
E= cM/4 = {M^2 \over {72 \pi}}.
\ee

\section*{Acknowledgements}
This work was supported in part by the DOE and the INFN. F. C. would like to
thank Darryl Holm for useful suggestions.

\section*{FIGURE CAPTIONS}

\noindent Fig. 1a  $  u_{v} $  with $n=.876$  and $u$ for
$M=1$ given by (61) as a function of x for the case: $p=0$,
$\alpha=1/2$, $l=3$.

\noindent Fig. 1b  $u- u_{v}$ as a function of x for $p=0$,
 $\alpha=1/2$, $l=3$.

\noindent Fig. 2a  $  u_{v} $ with $n=1.423$  and $u$ for $M=1$ given
by (64) as a function of x for the case: $p=2$, $\alpha=1/4$, $l=3$.

\noindent Fig. 2b  $u- u_{v}$ as a function of x for $p=0$,
 $\alpha=1/2$, $l=3$.

\noindent Fig. 3a  $  u_{v} $ with $n=1.155$  and $u$ for $M=1$  as a
function of x given by (68) for the case:  $p=1$, $\alpha=1/2$, $l=3$.

\noindent Fig. 3b  $ u_{v}- u$ as a function of x
for $p=1$, $\alpha=1/2$, $l=3$.

\noindent Fig. 4a  $  u_{v} $ with $n=1.283$  and $u$ for $M=1$ given by (72)
as a
function of x for the case: $p=2$, $\alpha=3$, $l=4$.

\noindent Fig. 4b  $ u_{v}-u$ as a function of x for $p=2$, $\alpha=3$,
$l=4$.


\begin{thebibliography}{99}
\bibitem{RH}P.Rosenau and J.M.Hyman, Phys. Rev. Lett. 70,  564 (1993).

\bibitem{Das} A. Das, Integrable Models (World Scientific Lecture
Notes in Physics, Vol.30) 1989.

\bibitem{CSLS1}
 F.Cooper, H. Shepard, C. Lucheroni, and P. Sodano,
``Post-Gaussian Variational Method for the Nonlinear Schrodinger Equation:
Soliton Behavior and
Blowup,'' Physica D (to be published).

\bibitem {CSLS2} F.Cooper,C. Lucheroni, H. Shepard and P. Sodano, Phys.Lett. A
173,  33 (1993).

\bibitem{Dirac}
P. A. M. Dirac, Proc. Camb. Phil. Soc. 26,  376 (1930).

\bibitem{Jackiw-Kerman}
R. Jackiw and A. Kerman, Phys. Lett. 71A,  158 (1979).


\bibitem{Jackiw-Faddeev}
L. Faddeev and R. Jackiw, Phys. Rev. Lett. 60, 1692 (1988).


\end {thebibliography}
\end{document}